\documentclass[aps,twocolumn,showpacs]{revtex4}

\usepackage{amsmath,amssymb,amsfonts}
\usepackage{graphics,graphicx}
\usepackage{epsfig,wrapfig}
\usepackage{color}

\begin{document}

\title{Compensating the Noise of a Communication Channel\\
via Asymmetric Encoding of Quantum Information}

\author{Marco Lucamarini$^{1}$}
\author{Giovanni Di Giuseppe$^{2}$}
\author{David Vitali$^{2}$}
\author{Paolo Tombesi$^{2}$}

\affiliation{\vspace{0.2cm}$^{1}$CNISM UdR, University of
Camerino, Via Madonna delle Carceri 9, 62032 Camerino (MC), Italy\\
\vspace{0.05cm}$^{2}$Physics Department, University of Camerino,
Via Madonna delle Carceri 9, 62032 Camerino (MC), Italy}

\date{\vspace{0.1cm}\today}

\begin{abstract}
\noindent An asymmetric preparation of the quantum states sent
through a noisy channel can enable a new way to monitor and
actively compensate the channel noise. The paradigm of such an
asymmetric treatment of quantum information is the Bennett 1992
protocol, in which the ratio between conclusive and inconclusive
counts is in direct connection with the channel noise. Using this
protocol as a guiding example, we show how to correct the phase
drift of a communication channel without using reference pulses,
interruptions of the quantum transmission or public data
exchanges.
\end{abstract}

\pacs{03.67.Dd, 03.65.Hk}

\maketitle


In recent years the field of Quantum Key Distribution
(QKD)~\cite{GRT+02} has reached such a high degree of technical
perfection that the emergence of unexplored directions in one of
its oldest protocols, the Bennett 1992~\cite{B92}, is somewhat
surprising. The Bennett 1992 (B92) protocol is based on only two
quantum states, $|\varphi_0\rangle$ and $|\varphi_1\rangle$,
strictly nonorthogonal, to which are associated the two values of
the logical bit communicated by the transmitter (Alice) to the
receiver (Bob)~\cite{B92}. This simple structure is one of the
main advantages of the B92 protocol, which also features
unconditional security~\cite{B92a} and suitability for
long-distance communications~\cite{B92b,LDT09}.

There is however another peculiarity of the B92 protocol which has
not been considered so far and yet it represents a useful resource
in the field of quantum information, i.e. the \textit{asymmetric
distribution} of its signal states, $|\varphi_0\rangle$ and
$|\varphi_1\rangle$.
In the left inset of Fig.~1 we depicted these states as two arrows
lying on the equator of the Poincar\'{e} sphere~\cite{BW99} at an
angle $\theta$ from the horizontal axis (noiseless case). In this
representation, orthogonal states are associated to antiparallel
arrows, to allow for a one-to-one map between a generic quantum
state and a single point of the Poincar\'{e} sphere~\cite{BW99}.
A symmetric distribution of the states would be if
$|\varphi_0\rangle$ and $|\varphi_1\rangle$ were lying in opposite
directions respect to the origin of the axes, antiparallel to each
other. But then they would be orthogonal, and this can never
happen in the B92 protocol. Hence the distribution of the signal
states is necessarily asymmetric for this protocol.

It turns out that this kind of asymmetry has its own advantages if
adequately exploited. In the B92 protocol Bob performs a
measurement of the incoming states and divides the results into
two main groups, labeled as \textit{conclusive} and
\textit{inconclusive}. The ratio of these two sets of data is
directly related to the noise of the channel thus allowing Bob to
quantify and correct it.

In the following, we apply this method to correct the
\textit{phase drift} of a communication channel. The method can be
easily generalized to include other sources of noise, e.g. that
due to the birefringence of an optical fiber affecting the
polarization of a light pulse. The phase drift model is sketched
in Fig.~1: in the left illustration the signal states reach the
receiver without being affected by the noise; on the contrary, in
the central and right illustrations, the states are rotated by an
angle $\varepsilon$ about the central axis and Bob's measurement
will consequently contain some errors. We call $\varepsilon$ the
\textit{misalignment angle}.
\begin{figure}[t]
  \includegraphics[width=8.5cm]{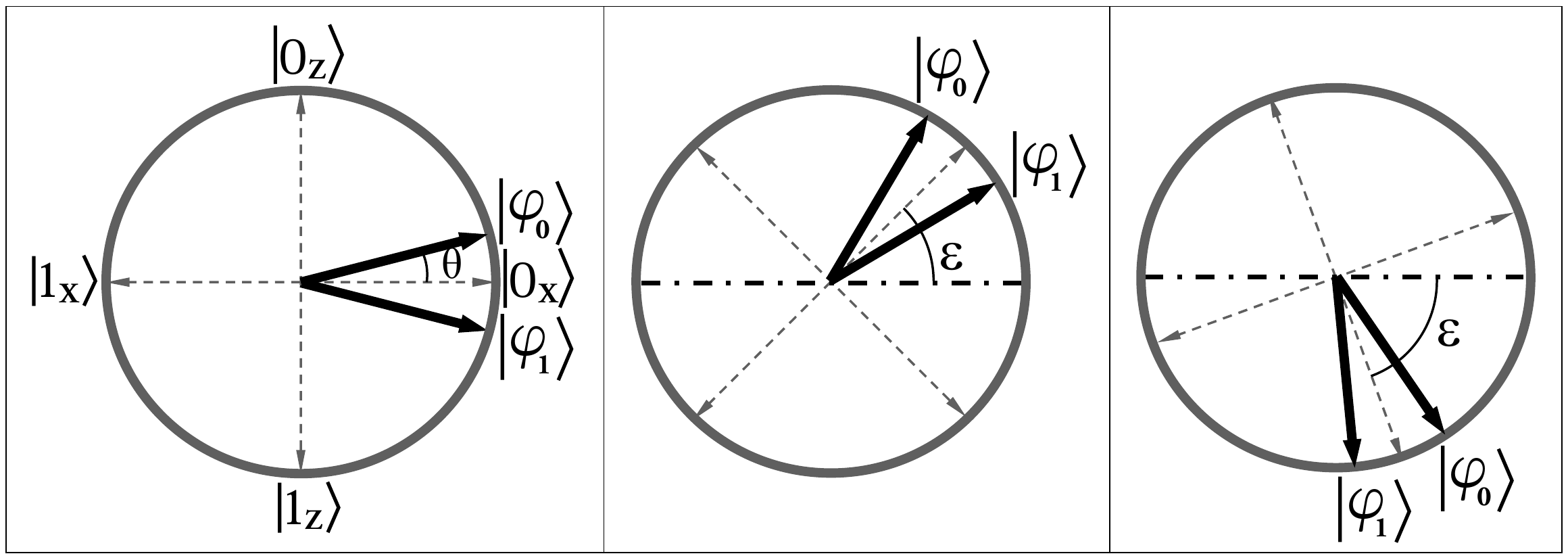}\\
  \caption{ States of the B92 protocol with a characteristic angle
  $\theta=\frac{\pi}{12}$~($15^{\circ}$). \underline{Left}:
  channel without phase-drift, $\varepsilon=0$.
  \underline{Center}:~~channel~with~a~phase-drift~of~$\varepsilon=\frac{\pi}{4}$~~($45^\circ$).
  \underline{Right}:~channel~with~a~phase-drift~of~$\varepsilon=-\frac{7}{18}\pi$~($-70^\circ$).
  The phase-drift rotates the reference system during the transfer
  of the quantum states from the transmitter to the receiver~\cite{BRS07}.
  }\label{fig:Fig1}
\end{figure}

Although the phase drift model can appear naive, it applies to a
vast number of QKD setups, precisely to all those in which the
information is encoded in the relative phase of a photon passing
through an unbalanced interferometer~\cite{B92,GRT+02}. To show
how relevant it is and to facilitate a comparison between our
proposal and existing solutions, we give in the next a brief
overview of the main techniques used to compensate the phase
drift.

One popular solution is to multiplex in the same channel the
quantum signals, e.g. single-photon pulses or attenuated laser
pulses, and the classical ones, e.g. intense
light-pulses~\cite{MT95,YS05}. In this way a technique already
employed in classical communications is adapted to the quantum
realm, but carries a few drawbacks though. First, it has been
recently shown that nonlinear effects due to the propagation of
intense light pulses in optical fibers can generate noise in the
sensitive single-photon detectors, thus limiting in practice the
maximum distance of a fiber-based QKD~\cite{SZT05}. Second, the
co-existence in the same channel of dim and bright pulses makes it
hard to control the intensity of the incoming light, and this can
be exploited by an eavesdropper to hack the QKD
system~\cite{Mak09}. Finally this technique requires additional
hardware to be implemented.

A similar analysis holds for those systems which use a two-way
configuration to enable a passive compensation of the
phase-drift~\cite{ZGG+97}. Indeed the pulse sent in the forward
direction is necessarily intense, thus causing Rayleigh
backscattering and opening the door to an interrogation
attack~\cite{VMH01}. Moreover, this technique cannot be employed
with a single-photon source, even if some attempts have been done
in this direction~\cite{LM05}.

There also exist solutions which employ quantum signals only. In
one case~\cite{MBH04} the quantum transmission is interrupted and
a sequence of quantum signals with a fixed phase value acting as a
reference is sent by the transmitter along the channel until the
receiver announces that the alignment has been completed . In
another case~\cite{EPS+03} the quantum signals with a fixed phase
value are inserted in the communication using a different
wavelength, like in the classical frequency-multiplexing
technique. Both of these solutions have some disadvantages. In the
former, the interruption of the quantum transmission represents an
idle cycle from which no secure bit can be distilled; moreover
this technique is not adaptive, i.e. the system cannot adapt the
interruption frequency to the real noise present on the channel.
In the latter, the presence of an extra wavelength implies that
single-photon detectors and electronics must deal with two
wavelengths rather than one, thus increasing the complexity of the
setup and reducing the key generation rate due to the fewer
detector's windows available for the signal states.

Our solution represents an alternative way to control the channel
noise at the quantum level without any of the above drawbacks. The
necessary resources for the control are borrowed from the
asymmetry of the same signals used for the very QKD process. So
there is no need to multiplex quantum signal with suitably
tailored bright or reference pulses, or to interrupt the quantum
communication at all.

To explain in detail how the compensation technique works it is
useful to recall the B92 protocol encoding-decoding
mechanism~\cite{B92}. Let us write explicitly the quantum states
of the protocol:
\begin{eqnarray}
    |\varphi_j\rangle &=& \beta |0_x\rangle + (-1)^j \alpha|1_x\rangle, \label{eq:b92states01} \\
    |\overline{\varphi}_j\rangle &=& \alpha |0_x\rangle - (-1)^j \beta|1_x\rangle.\label{eq:b92states02}
\end{eqnarray}
In the above equation $|\varphi_j\rangle$ are the signal states of
the protocol and $|\overline{\varphi}_j\rangle$ are the states
orthogonal to them, with $j=\{0,1\}$; $|0_x\rangle$ and
$|1_x\rangle$ are the eigenstates of the Pauli operator
$\textbf{X}$; $\beta =\sqrt{1-\alpha^2} = \cos(\theta/2)$ and
$\theta$ belongs to the open interval $(0,\pi/2)$.

To start the communication, Alice chooses at random the value of
the bit $j$, encodes it in the corresponding state
$|\varphi_j\rangle$ and transmits it to Bob.
The resulting density matrix $\rho$ prepared by Alice is then:
\begin{eqnarray}\label{eq:B92densitymatrix}
\nonumber  \rho &=& \left(|\varphi_0\rangle \langle \varphi_0| +
|\varphi_1\rangle \langle
    \varphi_1|\right)/2 \\
   &=& \beta^2 |0_x\rangle \langle 0_x|+ \alpha^2 |1_x\rangle \langle 1_x|.
\end{eqnarray}
It can be easily verified that the above density matrix is
\textit{asymmetric} because it is not proportional to the identity
operator in the 2-dimension Hilbert space, $\textbf{I} =
|0_x\rangle \langle 0_x|+ |1_x\rangle \langle 1_x|$. This is a
consequence of the strict nonorthogonality of the B92 protocol
states.
To decode the information, Bob measures the incoming states in the
basis $\textbf{B}_k = \{ |\varphi_k\rangle,
|\overline{\varphi}_k\rangle \}$, $k=\{0,1\}$.
Upon obtaining the state $|\overline{\varphi}_k\rangle$, Bob
decodes Alice's bit as $j=k\oplus1$ (the symbol $\oplus$ means
``addition modulo 2'') and labels the result as
\textit{conclusive}; on the contrary, upon obtaining the state
$|\varphi_k\rangle$, Bob is not able to decode Alice's bit
deterministically, and simply labels the result as
\textit{inconclusive}. For example, if Bob detects the state
$|\overline{\varphi}_0\rangle$, then he can say with certainty
that the prepared state was $|\varphi_{1}\rangle$, because
$|\varphi_{0}\rangle$ is orthogonal to the detected state. The
same is true for the detection of $|\overline{\varphi}_1\rangle$,
which indicates that $|\varphi_{0}\rangle$ was prepared. These are
examples of conclusive results. On the contrary, if Bob detects
the state $|\varphi_{0}\rangle$, he will not be able to
deterministically infer Alice's preparation, because that state
has a nonzero probability to come either from
$|\varphi_{0}\rangle$ or from $|\varphi_{1}\rangle$. Hence this is
an example of inconclusive result.

The heart of our technique is that the asymmetric density matrix
of Eq.\eqref{eq:B92densitymatrix} produces different amounts of
conclusive and inconclusive results in Bob's measure. In
particular
the ratio between inconclusive and conclusive counts is a function
of the angle $\theta$, known to the users, and of the noise. Bob
can estimate it during the quantum transmission thus obtaining, in
real time, information about the noise, useful to eventually
compensate it.


To establish the connection with the noise, let us consider a
communication channel affected by the phase-drift model of noise
(Fig.~1). In this case the density matrix seen by Bob is no more
the one given by Eq.\eqref{eq:B92densitymatrix} but rather a new
matrix $\widetilde{\rho}$ composed by the noise-affected quantum
states $|\widetilde{\varphi}_0\rangle$ and
$|\widetilde{\varphi}_1\rangle$. These states can be obtained by
rotating those of Eq.\eqref{eq:b92states01} through the operator
$\textbf{U}_\varepsilon=\exp\left(i\frac{\varepsilon}{2}\textbf{Y}\right)$,
with $\textbf{Y}$ the usual Pauli operator and $\varepsilon$ the
misalignment angle:
\begin{eqnarray}
\left\vert\widetilde{\varphi}_{0}\right\rangle&=&
\cos[(\theta+\varepsilon)/2]\left\vert
0_{x}\right\rangle+\sin[(\theta+\varepsilon)/2]\left\vert
1_{x}\right\rangle,\label{eq:b92noisystates01}\\
\left\vert\widetilde{\varphi}_{1}\right\rangle&=&
\cos[(\theta-\varepsilon)/2]\left\vert
0_{x}\right\rangle-\sin[(\theta-\varepsilon)/2] \left\vert
1_{x}\right\rangle.\label{eq:b92noisystates02}
\end{eqnarray}
%
%
From the density matrix $\widetilde{\rho}$ is then possible to
calculate the various probabilities associated to Bob's
measurement. To make our description more realistic we introduce
the quantity $\eta$, which is the probability to detect a
single-photon state, i.e. a state different from a vacuum or a
multi-photon state. It can be thought as a sort of \textit{total
transmission} of the QKD setup, including the transmission of the
communication channel, $\eta_{C}$, the efficiency of Bob's
detectors, $\eta_{B}$, and the probability of double clicks in
Bob's detectors~\cite{B92a}. With this in mind, we write down the
probability that Bob gets an inconclusive outcome,
$P_{k}^{inc}=\eta\left\langle\varphi_{k}\right\vert\widetilde{\rho}\left\vert\varphi_{k}\right\rangle$,
or a conclusive outcome,
$P_{k}^{con}=\eta\left\langle\overline{\varphi}_{k}\right\vert\widetilde{\rho}\left\vert\overline{\varphi}_{k}\right\rangle$,
when he measures in the basis $\textbf{B}_k$:
\begin{eqnarray}
  P_{k}^{inc} &=&
  \eta\{2+\cos\varepsilon+\cos[2\theta-(-1)^{k}\varepsilon]\}/4 \label{eq:Pkinc}\\
  P_{k}^{con} &=&
  \eta\{2-\cos\varepsilon-\cos[2\theta-(-1)^{k}\varepsilon]\}/4. \label{eq:Pkcon}
\end{eqnarray}
The ratio $R_k$ of the two above probabilities is independent of
$\eta$ and is a crucial quantity, called \textit{control
function}:
\begin{equation}\label{eq:Rk}
    R_k(\theta,\varepsilon)=\frac{2+\cos\varepsilon+\cos[2\theta-(-1)^{k}\varepsilon]}
    {2-\cos\varepsilon-\cos[2\theta-(-1)^{k}\varepsilon]}.
\end{equation}
A few examples of $R_k$ are plotted in Fig.~2; the parameters used
to draw the curves are $k=\{0,1\}$ and
$\theta=\{\frac{5\pi}{18},\frac{\pi}{3},\frac{4\pi}{9}\}$. Among
these curves, some are better than others to drive the
noise-compensation mechanism. If the absolute value of $R_k$ is
too small, like for the curves with $\theta=\frac{4\pi}{9}$, the
system is less responsive i.e. big changes of $\varepsilon$ cause
small changes of $R_k$. In this case the risk is that the
misalignment angle becomes too large before Bob becomes aware of
it and applies the correction mechanism. On the contrary, the
higher the $R_k$ the fewer the conclusive counts registered by
Bob. This results in larger fluctuations in the estimation of
$R_k$, thus worsening the compensation mechanism, as it happens
for example to the curves featuring $\theta=\frac{5\pi}{18}$ in
Fig.~2. So the best option is to choose an intermediate value of
$\theta$ that at the same time provides Bob with a good statistics
and a responsive control. By consequence we choose
$\theta=\frac{\pi}{3}$ ($60^{\circ}$) and plot in Fig.~2 the
corresponding control functions together with their tangents in
the zero-noise point. Such a value of $\theta$ is also interesting
because it allows to merge in a single protocol the present
technique and that described in~\cite{LDT09}, that is an efficient
long-distance version of the B92 protocol featuring an optimal
$\theta$ of about $0.3\pi$ (precisely,
$55.4^{\circ}$~\cite{LDT09}).

When the noise is small, the control functions of Eq.\eqref{eq:Rk}
can be well approximated by their tangents in the zero-noise
point; furthermore they are monotone, so only one control
function, either $R_0$ or $R_1$, is sufficient to provide a
reliable estimation of $\varepsilon$. This makes the feedback
response very fast and we refer to this situation as to a ``fast
feedback''. On the contrary, when the noise is large and falls
outside the monotonicity range of the control functions, Bob must
use both the control functions to estimate unambiguously the
misalignment angle $\varepsilon$ in the open interval
$(-\pi,\pi)$. This procedure is intrinsically slower than the
previous one, so we term it ``slow feedback''. From a practical
point of view, both the options are helpful. For example, at the
beginning of a communication the users' apparatuses are completely
misaligned and $\varepsilon$ can be quite large; Bob will then use
the slow feedback to get a first estimation of $\varepsilon$ and
compensate it; after that, when the boxes are nearly aligned, Bob
will use the fast feedback to improve the alignment and maintain
it during the remaining quantum transmission.
\begin{figure}
  \includegraphics[width=8.5cm]{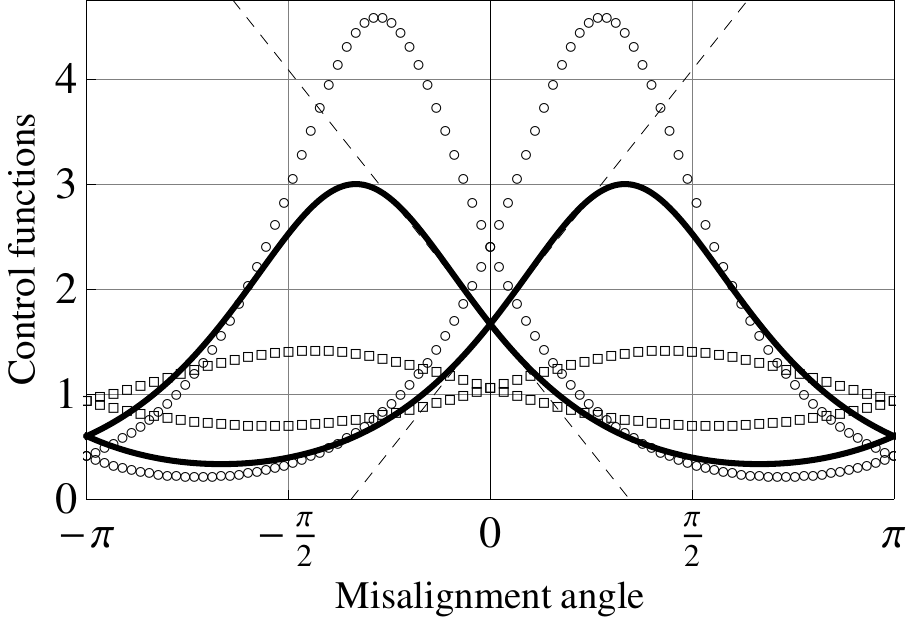}\\
  \caption{Examples of control functions
$R_k(\theta,\varepsilon)$ versus the misalignment angle
$\varepsilon$. The $R_0$ curves ($R_1$ curves) reach their maximum
in the right-part (left-part) of the figure. The figure shows that
a given misalignment angle causes a correspondent value in the
control functions. Bob can exploit this correspondence to measure
the control functions and from that ascertain the value of the
misalignment angle. \newline The $R_k(\frac{\pi}{3},\varepsilon)$
are drawn with solid lines; the $R_k(\frac{5\pi}{18},\varepsilon)$
with empty circles; the $R_k(\frac{4\pi}{9},\varepsilon)$ with
empty squares. The tangents to $R_k(\frac{\pi}{3},\varepsilon)$ in
the zero-noise point are in dashed lines.}\label{fig:fig02}
\end{figure}
%


At this point let us mention a possible experimental
implementation of the mechanism used to correct the noise. The QKD
layout we are interested in is the one based on the relative-phase
degree of freedom, in a one-way configuration. It is constituted
by two identical interferometers placed in Alice's and Bob's
stations~\cite{B92}. Each of the interferometers features two
unequal paths which provide the traveling pulses with a different
amount of optical phase. The phase difference can be easily
modulated by the users, who are then able to encode information in
this way. In particular, the output ports of the receiver's
interferometer are usually connected with two detectors that
determine the conclusiveness or inconclusiveness of a certain
result.

To correct the noise, Bob executes his measurement for a while,
registering all the results in a computer memory. Then, when a
sufficient number of occurrences is available, Bob estimates the
control functions of Eq.\eqref{eq:Rk} and obtains a value of the
misalignment angle $\varepsilon$. The last thing Bob has to do is
to recalibrate his phase-modulator with the obtained value to
re-establish the correct alignment of his apparatus with Alice's.
The read-and-write procedure just described is performed by Bob
in-course-of-action, and can be easily implemented by an
electronic feedback loop in which the estimated value of
$R_k(\theta,\varepsilon)$ is the \textit{input}, the fixed value
of $R_k(\theta,0)$ is the \textit{setpoint}, and a function of
these two quantities
is the \textit{output}.


We have carried out numerical simulations of such an active
feedback loop in the case of $\theta=\frac{\pi}{3}$ and the
results are reported in Fig.~3.
\begin{figure}
  \includegraphics[width=8.5cm]{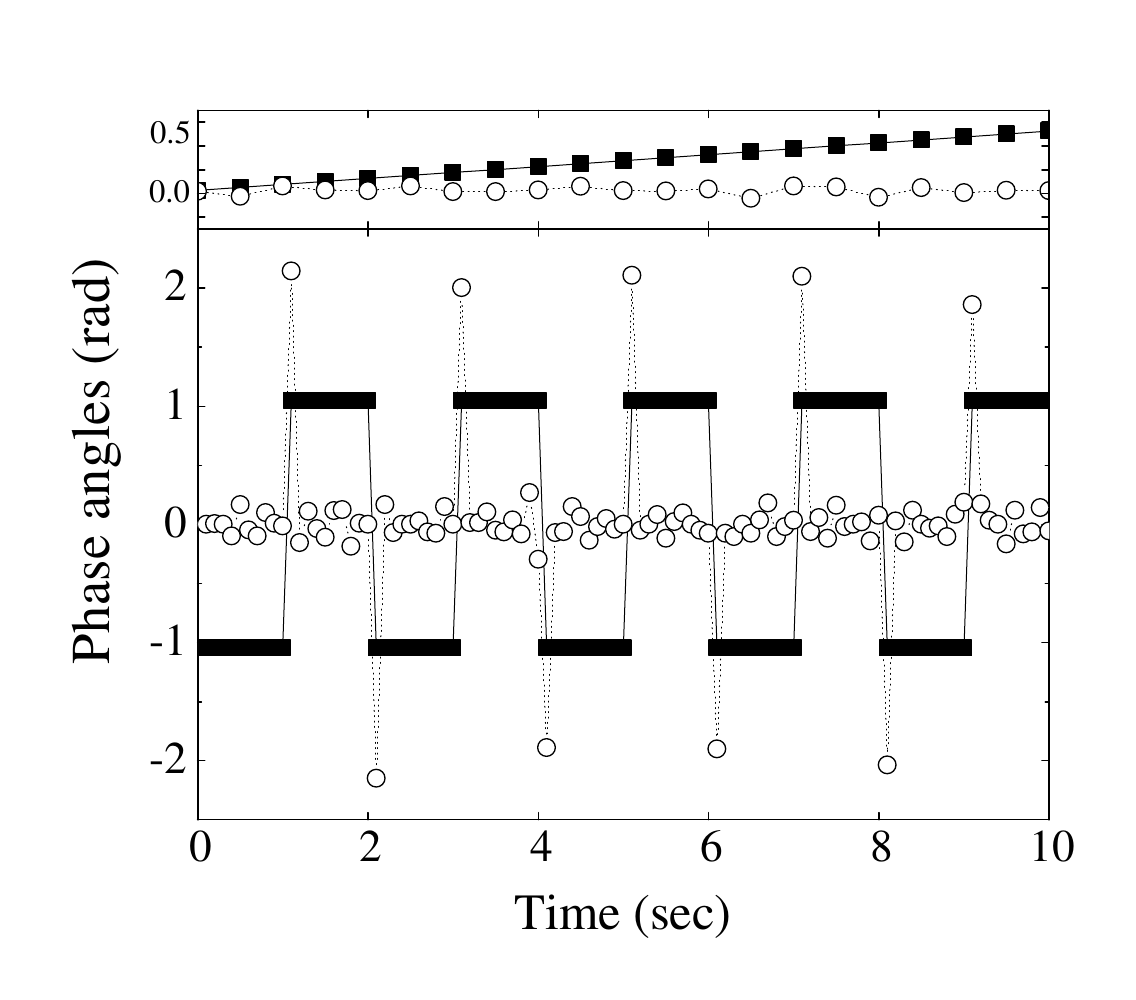}\\
  \caption{Noise patterns and feedback responses versus time.
  \underline{Top}: linear noise with a rate of $0.05$~rad/sec (filled squares)
  and response after fast feedback is applied with a statistics of
  $5\times10^{3}$~events per point (empty circles). The mean angle
  after feedback is $\varepsilon_{fb}^{fast}=(0.023 \pm 0.031)~\textrm{rad}$.
  \underline{Bottom}: step noise with amplitude 2~rad (filled squares) and response
  after slow feedback is applied with a statistics of $10^{3}$~events
  per point (empty circles). Other parameters used are (see text):
  $\theta=\frac{\pi}{3}$, $f=2$~MHz, $\mu=0.5$, $\eta_B=10\%$, $\eta_C=0.1$.}\label{fig:fig03}
\end{figure}
In the upper part of the figure we considered a linear increase in
time of the misalignment angle, with a rate equal to
$0.05$~rad/sec. This value is easily attainable with some care in
the shielding process of the users' interferometers against
external thermal fluctuations. Phase drift values reported in the
literature are well below our threshold, e.g. $0.033$~rad/sec
in~\cite{MBH04} or $0.0086$~rad/sec in~\cite{EPS+03}. We have also
assumed a trigger rate $f=2$~MHz, an average photon number per
pulse $\mu=0.5$, a detector efficiency $\eta_{B}=10\%$, a channel
transmittance $\eta_{C}=0.1$, equal to about $50$ km of a standard
optical fiber in the third Telecom window. Multiplied together,
these values provide Bob with a number of non-vacuum events per
second of the order of $10^{4}$. Let us note that this represents
a wide underestimation of the current QKD performances (see
e.g.~\cite{YDD+08}). Each point of the plotted curves is the
result of an average of $5\times10^{3}$ acquisitions, obtained in
practice by applying a feedback kick every half a second.
Following the approach in~\cite{EPS+03}, we have empirically
verified the optimality of this value: if larger, the phase-drift
would become too big in the between of two consecutive feedback
kicks; if smaller, the statistical error pertaining to Bob's
measurement would increase considerably.  The empty circles in the
upper part of Fig.~3 show the response of the \textit{fast}
feedback loop, i.e. the one in which a linear approximation of the
control functions is adopted. In particular we used the
first-order expansion of $R_0( \frac{\pi}{3},\varepsilon)$ in
$\varepsilon=0$ to drive the feedback. With feedback, the mean
misalignment angle of the communication channel reduces to:
\begin{equation}\label{eq:comp-angle-fast}
    \varepsilon_{fb}^{fast}=(0.023 \pm 0.031)~\textrm{rad}.
\end{equation}
The slight bias towards positive values is due to the monotone
increase of the phase-drift. The standard deviation is well below
the amount of phase-drift accumulated every second on the channel.

In the lower part of Fig.~3 we have considered a phase-drift in
the form of a step-function of amplitude equal to $2$ rad, to
study the response of the \textit{slow} feedback loop in the case
of a large phase-drift. In this case Bob exploits both
$R_0(\frac{\pi}{3},\varepsilon)$ and
$R_1(\frac{\pi}{3},\varepsilon)$ to estimate the misalignment
angle. After acquiring $10^{3}$ events he evaluates the mean
values of $R_0$ and $R_1$, let us term them $R_0^*$ and $R_1^*$.
Then he numerically finds the point $\varepsilon^{*}$ that
minimizes the quantity $|R_0(\frac{\pi}{3},\varepsilon)-R_0^*|$
$+$ $|R_1(\frac{\pi}{3},\varepsilon)-R_1^*|$. He finally uses the
value of $\varepsilon^*$ to compensate the phase-drift. The empty
circles of the figure again represent the value of the
misalignment angle after the feedback has been applied. It can be
seen that in correspondence of the noise steps the angle seen by
Bob undergoes strong jumps, but the system recovers immediately
after the jump. If one ignores the jumps, the mean misalignment
angle is:
\begin{equation}\label{eq:comp-angle-slow}
    \varepsilon_{fb}^{slow}=(-0.002 \pm 0.098)~\textrm{rad}.
\end{equation}
Compared to Eq.\eqref{eq:comp-angle-fast}, the average value is
nearly unbiased, following the unbiased behavior of the noise,
while the standard deviation is more than three times bigger due
to the reduced statistics adopted for this kind of feedback.
It can be nice to note that the step-noise pattern of Fig.3 could
be employed by the users to create an \textit{additional
communication channel} between them. More explicitly, Alice could
purposely feed into her phase-modulator a large phase value in
order to send some sort of message to Bob, like e.g. a string of
bits useful to identify a certain part of the quantum
transmission.


As a last point of our work we want to establish whether the
results of Eqs.~\eqref{eq:comp-angle-fast} and
\eqref{eq:comp-angle-slow} are good enough for a practical
implementation of the B92 protocol. In particular we calculate the
maximum misalignment angle for which the secure gain of the B92
protocol is still positive. The secure gain of a protocol is the
ratio between the number of secure bits distilled and the number
of qubits prepared by Alice. For the B92 protocol it is defined as
$G=\Lambda_{con}[1-h(\frac{\Lambda_{bit}}{\Lambda_{con}})
-h(\frac{\overline{\Lambda}_{ph}}{\Lambda_{con}})]$, with $h$ the
Shannon entropy, $\Lambda_{con}$ the conclusive-count rate,
$\Lambda_{bit}$ the bit-error rate and $\overline{\Lambda}_{ph}$
the upper bound to the phase-error rate obtained from
$\Lambda_{bit}$ and $\Lambda_{con}$ through an optimization
algorithm~\cite{B92a,LDT09}.

The bit-error rate is given by the probability that Bob finds the
state $\left\vert \overline{\varphi}_{0}\right\rangle $
($\left\vert \overline{\varphi}_{1}\right\rangle $) when Alice
prepares the orthogonal state $\left\vert
\varphi_{0}\right\rangle$ ($\left\vert \varphi_{1}\right\rangle$).
For the present discussion it suffices to assume that the
phase-drift is the only source of errors; in reality there are
also the detector dark counts and the unavoidable experimental
imperfections. By using the noisy states of
Eqs.~\eqref{eq:b92noisystates01} and~\eqref{eq:b92noisystates02},
and assuming a preparation of the bit value 0 (the same holds for
the bit value 1), we obtain the following bit-error rate:
\begin{eqnarray}
\nonumber  \Lambda_{bit} &=& \eta(\left\vert \left\langle
\widetilde{\varphi}_{0}|\overline{\varphi}_{0}\right\rangle
\right\vert ^{2})/2 \\
&=& \eta\left( 1-\cos\varepsilon\right)/4 .\label{eq:Lambda-bit}
\end{eqnarray}
This expression is independent of $\theta$ and correctly vanishes
when $\varepsilon$ tends to zero. The coefficient $\eta$ takes
into account the vacuum or multi-photon counts by Bob, while the
factor $1/2$ is the probability to guess the right basis to detect
the error.


In a similar way, the conclusive-count rate is given by the
probability that Bob obtains a conclusive result, and can be
evaluated from the mean value of $P_0^{con}$ and $P_1^{con}$ in
Eq.~\eqref{eq:Pkcon}:
\begin{eqnarray}
\nonumber \Lambda_{con} &=& (P_0^{con}+P_1^{inc})/2 \\
&=& \eta(1-\cos\varepsilon\cos^{2}\theta)/2.\label{eq:Lambda-con}
\end{eqnarray}
This time, there is a dependance on $\theta$. After fixing
$\theta=\frac{\pi}{3}$ and using Eqs.~\eqref{eq:Lambda-bit} and
\eqref{eq:Lambda-con}, we have numerically found a positive gain
for the B92 protocol until
\begin{equation}\label{noisemax}
    |\varepsilon|<0.27646~\textrm{rad}.
\end{equation}
The value of $\varepsilon$ given by the fast feedback,
Eq.~\eqref{eq:comp-angle-fast}, is much smaller than the above
threshold. Hence our proposal appears feasible, especially if one
considers that we made conservative assumptions on the QKD
parameters and that less conservative assumptions would lead to
better values in Eqs.~\eqref{eq:comp-angle-fast} and
\eqref{eq:comp-angle-slow}.

Let us point out that Eqs.~\eqref{eq:Lambda-bit} and
\eqref{eq:Lambda-con} establish a direct dependance of the
bit-error rate and conclusive-count rate on the misalignment angle
$\varepsilon$. This allows Bob to estimate $\Lambda_{bit}$ and
$\Lambda_{con}$ \textit{without communicating with Alice}. It
suffices that Bob gets an estimate of $\varepsilon$ from his data
and substitutes it into the given equations~\cite{note2}. This is
a further peculiarity of the B92 protocol that remained unnoticed
so far. It can considerably increase the practicality of the
protocol by reducing the classical communication necessary to
distill the final key.


In conclusion, we introduced a novel scheme to detect and correct
the phase-drift of a communication channel based on some
characteristics of the B92 protocol not considered so far. The
scheme features a few remarkable properties: it is entirely
quantum, it can be executed real-time without interrupting the
communication, it allows to estimate the bit-error rate without a
bidirectional communication, it creates additional communication
channels for the users. This highly increases the practicality of
the B92 protocol, often considered unsuitable for real-world
implementations. Furthermore, the fully quantum nature of the
scheme on one side reduces the noise due to the propagation of
high-intensity light pulses in a nonlinear medium, on the other
makes it conceivable the construction of networks and devices
working entirely at the quantum level, thus preventing several
hacking strategies available to Eve. The asymmetry-based
correction mechanism can be extended to the polarization degree of
freedom and can play a role in the entanglement distribution
problem.


\begin{thebibliography}{99}

\bibitem{GRT+02} N. Gisin, G. Ribordy, W. Tittel and H. Zbinden,
Rev. Mod. Phys. \textbf{74}, 145 (2002).

\bibitem{B92} C. H. Bennett, Phys. Rev. Lett. 68, 3121 (1992).

\bibitem{B92a} K. Tamaki, M. Koashi, and N. Imoto, Phys. Rev. Lett. \textbf{90},
167904 (2003); K. Tamaki and N. L\"{u}tkenhaus, Phys. Rev. A
\textbf{69}, 032316 (2004).

\bibitem{B92b} M. Koashi, Phys. Rev. Lett. \textbf{93}, 120501 (2004);
K. Tamaki, N. L\"{u}tkenhaus, M. Koashi, and J. Batuwantudawe,
Phys. Rev. A \textbf{80}, 032302 (2009); K. Tamaki, \textit{ibid.}
\textbf{77}, 032341 (2008).

\bibitem{LDT09} M. Lucamarini, G. Di Giuseppe, and K. Tamaki, Phys. Rev. A \textbf{80}, 032327
(2009).

\bibitem{BW99} M. Born and E. Wolf, \textit{Principles of
Optics}, 7th (expanded) edition, Cambridge University Press
(1999).

\bibitem{BRS07} S. D. Bartlett, T. Rudolph, and R. W. Spekkens, Rev. Mod. Phys.
\textbf{79}, 555 (2007).

\bibitem{MT95} C. Marand and P. Townsend, Opt. Lett. \textbf{20}, 1695 (1995).

\bibitem{YS05} Z. L. Yuan and A. J. Shields, Opt. Expr. \textbf{13}, 660
(2005).

\bibitem{SZT05} D. Subacius and A. Zavriyev and A. Trifonov, Appl. Phys. Lett. \textbf{86}, 011103
(2005).

\bibitem{Mak09} V. Makarov, New J. Phys. \textbf{11}, 065003
(2009).

\bibitem{ZGG+97} H. Zbinden, J. D. Gautier, N. Gisin, B. Huttner, A. Muller, and W.
Tittel, Electron. Lett. \textbf{33}, 586 (1997).

\bibitem{VMH01} A. Vakhitov, V. Makarov, and D. R. Hjelme, J. Mod. Opt. \textbf{48}, 2023
(2001).

\bibitem{LM05} M. Lucamarini and S. Mancini, Phys. Rev. Lett. \textbf{94}, 140501 (2005);
A. Cer\`{e}, M. Lucamarini, G. Di~Giuseppe, and P. Tombesi,
\textit{ibid.} \textbf{96}, 200501 (2006); R. Kumar, M.
Lucamarini, G. Di~Giuseppe, R. Natali, G. Mancini, and P. Tombesi,
Phys. Rev. A \textbf{77}, 022304 (2008).

\bibitem{MBH04} V. Makarov, A. Brylevski, and D. R. Hjelme, Appl. Opt. \textbf{43}, 4385
(2004).

\bibitem{EPS+03} B. B. Elliott, O. Pikalo, J. Schlafer, and G. Troxel,
Proc. SPIE \textbf{5105}, 26 (2003).

\bibitem{YDD+08} Z. L. Yuan, A. R. Dixon, J. F. Dynes, A. W. Sharpe, and A. J.
Shields, Appl. Phys. Lett. \textbf{92}, 201104 (2008).

\bibitem{note2} Since Bob knows the expected number of single-photon events, he
can easily appraise $\eta$ from the single-photon events
effectively registered by his measuring device.

\end{thebibliography}
\end{document}